\documentclass{article}
\pdfoutput=1
\usepackage{spconf,amsmath,graphicx}
\usepackage{amsmath,amsthm,amsfonts,amssymb,mathtools,booktabs}
\usepackage[bookmarks=false]{hyperref}

\title{E3 TTS: Easy End-to-End Diffusion-based Text to Speech}
\name{Yuan Gao, Nobuyuki Morioka, Yu Zhang, Nanxin Chen}
\address{Google\\
\small \texttt{\{gaoyua,nmorioka,ngyuzh,nanxinchen\}@google.com}
}

\copyrightnotice{979-8-3503-0689-7/23/\$31.00~\copyright2023 IEEE}
\begin{document}
\maketitle
\begin{abstract}

We propose \textbf{E}asy \textbf{E}nd-to-\textbf{E}nd Diffusion-based Text to Speech, a simple and efficient end-to-end text-to-speech model based on diffusion.
E3 TTS directly takes plain text as input and generates an audio waveform through an iterative refinement process.
Unlike many prior work, E3 TTS does not rely on any intermediate representations like spectrogram features or alignment information.
Instead, E3 TTS models the temporal structure of the waveform through the diffusion process.
Without relying on additional conditioning information, E3 TTS could support flexible latent structure within the given audio.
This enables E3 TTS to be easily adapted for zero-shot tasks such as editing without any additional training.
Experiments show that E3 TTS can generate high-fidelity audio, approaching the performance of a state-of-the-art neural TTS system.
Audio samples are available at \url{https://e3tts.github.io}.
\end{abstract}
\section{Introduction}
\label{sec:intro}
Diffusion models \cite{ho2020denoising, sohldickstein2015deep, song2021scorebased} have demonstrated great performance on a variety of generation tasks, including image~\cite{saharia2022photorealistic,ramesh2022hierarchical, hoogeboom2023simple} and audio generation~\cite{chen2020wavegrad,kong2020diffwave}. Diffusion models work by gradually removing noise from a latent representation of the data until it becomes indistinguishable from real data. Text-to-speech (TTS) systems that use diffusion models have been shown to produce high-fidelity speech that is comparable with state-of-the-art systems~\cite{chen2020wavegrad, kong2020diffwave, chen2021wavegrad, liu2023diffvoice}.

Most prior work in this area has relied on a two-stage generation process.
In the first stage, the generator model generates intermediate representations, typically audio tokens~\cite{wang2023neural, borsos2023audiolm} or spectrogram-based features~\cite{ren2019fastspeech, li2019neural, shen2020non, elias2021parallel,kim2022guidedtts,tae2022editts,10096285}, which are aligned with the waveform but in lower resolution.
In the second stage, a vocoder is introduced to predict the audio from the intermediate features.
Besides the two-stage process, most models take some extra neural model or statistical method to convert the text to some other input units~\cite{sproat2022boring}, such as phonemes or graphemes.

Even though a two-stage TTS pipeline can produce higher quality audio, it may also have other concerns, such as relying on the quality of intermediate features. Additionally, it is more complicated to deploy and set up for different situations.

End-to-end generation of audio from text is elusive, due to the difficulty of efficiently modeling strong temporal dependencies in the waveform. 
Sample-level autoregressive vocoders handle such dependencies by conditioning generation of each waveform sample on the whole history.
Due to their highly sequential nature, they are inefficient to sample from on modern parallel hardware.
Some previous work instead generates a sequence of non-overlapping fixed-length blocks autoregressively to speedup the generation~\cite{weiss2021wavetacotron}.
This speeds up the generation process by generating all samples within the block in parallel.

A different direction of prior work is to include alignment information during training.
The alignment information provides the mapping between each individual input unit, such as a phoneme, and output samples in the generated audio.
It is usually extracted using external alignment tools, which provide the start time and end time of each individual input unit.
FastSpeech 2~\cite{ren2021fastspeech} relies on such alignment or duration information and other properties such as energy and pitch to predict the audio.
One internal predictor is also trained for each individual property so the predicted results could be utilized during inference.
EATS~\cite{donahue2021endtoend} proposes to use a differentiable duration predictor and depends on the Dynamic Time Wraping (DTW) to make sure the prediction is aligned with the target audio.
This avoids the usage of external aligner but makes the training more complicated.

In this paper, we propose a Easy End-to-End Text to Speech framework (E3 TTS) that only relies on diffusion to preserve temporal structure in waveform.  It directly takes text as input and uses a pretrained BERT model~\cite{kenton2019bert} to extract information from it.
It is followed by a UNet structure~\cite{ronneberger2015u} which predicts the audio by attending to the BERT representations.
The whole model is non-autoregressive and  directly outputs a waveform.
Our model achieves comparable results to the two-stage framework on proprietary dataset from experiments.

The paper is organized as follows.
Section \ref{sec:dur} gives a brief overview of different components used in prior works of TTS that could be optimized.
Section \ref{sec:method} introduces the proposed system which only includes a diffusion model taking BERT representations as input.
Section \ref{sec:experiment} starts with experiments on proprietary dataset, comparing with some previous work.
Section \ref{sec:application} reveals some applications that could be achieved with the proposed method.
Section \ref{sec:summary} summarizes the system and discusses some future work.

\section{Complexities of TTS}
\label{sec:dur}

Through a careful analysis of current text-to-speech (TTS) systems, we have identified several components that greatly increase the complexities of existing systems.

\subsection{Text Normalization}

One of the challenges in building a text-to-speech (TTS) system is the normalization of input text.
This is the process of converting text from its written form into a form that can be easily processed by the TTS system.
This can be a difficult task, as there are many different ways that text can be written~\cite{sproat2022boring}.
For example, the same word can be written in different ways, such as "color" and "colour". Additionally, text can contain abbreviations, acronyms, and other non-standard forms. 
A good TTS system must be able to handle all of these different variations in order to produce accurate and natural-sounding speech.

\subsection{Input Unit}

Even after text normalization, there can still be ambiguities in how to pronounce the same word in different contexts.
For example, {\sl record} has different pronunciations depending whether it is a noun or a verb.
This is why many TTS systems rely on verbalized forms, such as phonemes or prosodic features, instead of text directly.

{\bf Phonemes}: A phoneme is a unit of sound that is used to make up words. Many previous work~\cite{chen2020wavegrad, liu2023diffvoice} rely on phonemes as input. This can be useful for generating speech from languages that do not have a standard writing system.

{\bf Prosodic features}: Prosodic features are characteristics of speech, such as fundamental frequencies, durations, and energy. Some previous work~\cite{ren2019fastspeech, ren2021fastspeech} utilize prosodic features as input. This can be used to control the intonation and emphasis of the generated speech.

\subsection{Alignment Modeling}

Another challenge in building a TTS system is alignment modeling. This is the process of predicting the length of time that each phoneme in a word should be pronounced. This is important because it helps to ensure that the generated speech sounds natural and fluent. Alignment modeling can be a difficult task, as there are many factors that can affect the length of time that a phoneme is pronounced. For example, the position of a phoneme in a word can affect its duration. Additionally, the stress of a word can also affect the duration of its phonemes. A good TTS system must be able to model all of these factors in order to produce accurate and natural-sounding speech.

A typical approach for alignment modeling in end-to-end speech-to-text system is to rely on external aligner which provides the alignment information given transcript and audio~\cite{ren2019fastspeech,liu2023diffvoice}.
During model training, a duration predictor is learned to predict the information which could be used to estimate alignment for inference.
For duration predictor, Non-Attentive Tacotron framework \cite{shen2020non} managed to learn duration implicitly by employing the Variational Auto-Encoder. Glow-TTS \cite{10.5555/3495724.3496400} and Grad-TTS \cite{pmlr-v139-popov21a}  made use of Monotonic Alignment Search algorithm (an adoption of Viterbi training \cite{18626} finding the most likely hidden alignment between two sequences). Indeed, we actually solve the quality issue Grad-TTS mentioned in paper when they try to conduct end-to-end experiments.

\section{Method}
\label{sec:method}
\begin{figure*}[t!]
\includegraphics[width=1.0\textwidth]{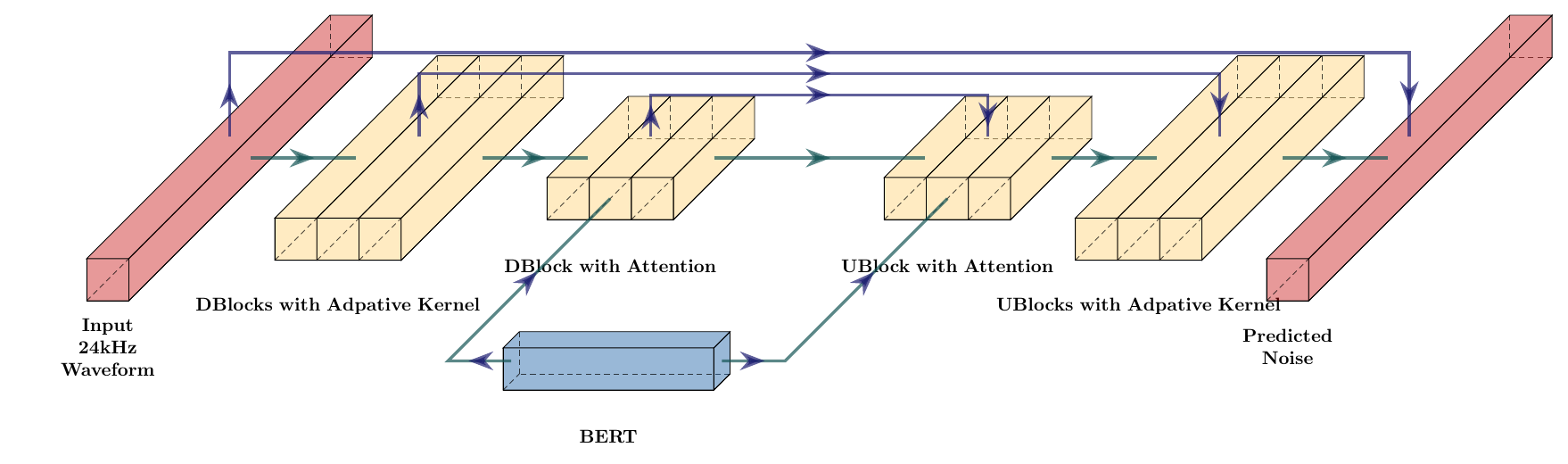}
\caption{
\label{fig:unet}
UNet Structure: DBlock for downsampling block, UBlock for upsampling block}
\end{figure*}

We propose our solution that addresses challenges presented in the last Section to make TTS systems more accessible to the wider community.
The proposed model includes two modules illustrated in Figure \ref{fig:unet}:
\begin{itemize}
    \item A pretrained BERT model extracts information from text.
    \item An diffusion UNet model attends to the BERT output and predicts the raw waveform by refining the noisy waveform iteratively.
\end{itemize}

\subsection{BERT model}
To take the advantage of the recent large language model development, we built our system based on the text representations which are given by a pretrained BERT model~\cite{kenton2019bert}.
The BERT model takes the subword as input and it does not rely on any other presentations of the speech such as phoneme, graphemes, in contrast to some previous work~\cite{shen2020non, elias2021parallel, li2019neural, ren2019fastspeech, ren2021fastspeech, weiss2021wavetacotron, wang2018style, skerry2018towards}.
This simplifies the process since one could rely on a pretrained text language model which could be trained on multiple languages with only text data available.

\subsection{Diffusion}
\label{sec:diffusion}
Our model is built based on prior work on score matching~\cite{song2021scorebased} and diffusion probabilistic models~\cite{ho2020denoising}.
In the case of TTS, the score function is defined as the gradient of the log conditional distribution $p(y \mid x)$ with respect to the output $y$ as
\begin{equation}
    s(y \mid x) = \nabla_y \log p(y \mid x)
\end{equation}
where $y$ is the waveform and $x$ is the conditioning signal.

Following previous work~\cite{chen2021wavegrad}, we adopt a special parameterization known as the diffusion model~\cite{ho2020denoising}. A score network $s(\tilde y \mid x, \bar \alpha)$ is trained to predict the scaled derivative by minimizing the distance between model prediction and ground truth $\epsilon$ as
\begin{align}
     \mathbb{E}_{\bar \alpha,\epsilon}\left[ \left\lVert \epsilon_\theta\left(\tilde y, x, \sqrt {\bar \alpha} \right) - \epsilon \right\rVert_2 \right]
\end{align}
where $\epsilon \sim \mathcal{N}(0, I)$ is the noise term introduced by applying the reparameterization trick, $\bar\alpha$ is the noise level and $\tilde y$ is sampled according to
\begin{equation}
    \tilde y = \sqrt{\bar \alpha}\, y_0 +\sqrt{1 - \bar\alpha}\, \epsilon
\end{equation}
During training, $\bar\alpha$'s are sampled from the intervals $\left[ \bar \alpha_n, \bar \alpha_{n+1} \right]$ based on a pre-defined linear schedule of $\beta$'s, according to:
\begin{equation}
\bar \alpha_n \coloneqq \prod_{s=1}^n (1 - \beta_s)
\end{equation}
In each iteration, the updated waveform is estimated following the following stochastic process
\begin{equation}
     y_{n-1} = \frac{1}{\sqrt{\alpha_n}}\left( y_n - \frac{\beta_n}{\sqrt{1-\bar\alpha_n}}\, \epsilon_\theta(y_n, x, \sqrt{\bar\alpha_n}) \right) + \sigma_n z
\end{equation}

In this work, to help convergence and to better scale $\epsilon$ loss's magnitude, we adopt a KL form loss.
The model also predicted the variance $\omega(\alpha)$ of the L2 loss according to timestep, and we use a KL loss form to adjust the weight of the loss from different sampled timestep.
\begin{align}
     \mathbb{E}_{\bar \alpha,\epsilon}\left[ \frac{1}{\omega({\bar\alpha)}} \left\lVert \epsilon_\theta\left(\tilde y, x, \sqrt {\bar \alpha} \right) - \epsilon \right\rVert_2  + ln(\omega({\bar\alpha}))\right]
\end{align}

\subsection{U-Net}

\begin{figure*}[h]
\centering
\includegraphics[width=0.9\textwidth]{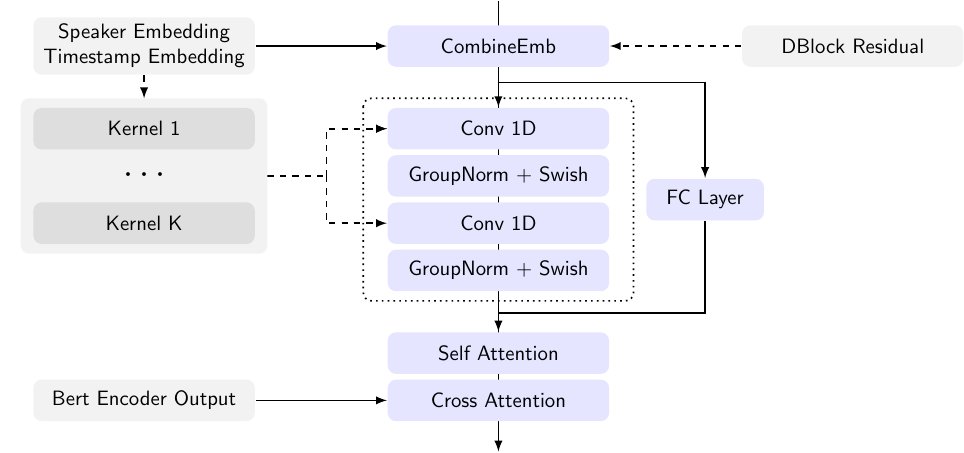}
\caption{
\label{fig:cnn_block}
UBlock/DBlock Struture: Adaptive Kernel and residual optional.}
\end{figure*}

We deploy a 1D U-Net, following the structure of \cite{saharia2022photorealistic}. The general model structure are shown in  Figure~\ref{fig:unet}, consists of a series of downsampling and upsampling blocks connected by residual. The detailed structure of each downsampling/upsampling block is shown in Figure~\ref{fig:cnn_block}.
Like the typical approach in autoregressive TTS  \cite{wang16e_interspeech,wang2017tacotron}, we adopt a \textbf{cross-attention} to extract information from BERT output in the top downsampling/upsampling blocks.
In the low downsampling/upsampling block, following \cite{kang2023scaling}, we use an adaptive softmax CNN kernel whose kernel is determined by timestep and speaker. In other layers, speaker and timestep embedding are joined using FiLM \cite{perez2017film}, which comprises a combined layer which predicts channel-wise scaling and bias. Inside each block, ublock and dblock the structure closely follow the structure described in \cite{saharia2022photorealistic}.

The downsampler finally refined the noise information (24kHz) to a sequence whose length is similar to the encoded BERT output.
This has proved important in practice to improve the quality. The upsampler finally predicts noise whose length is the same as the input waveform.

In training, we fixed the length of waveform to be 10.92 sec, and padding zero to the end of the waveform. When calculating loss, the padding part are less weighted. In practice, we weight each padding frame $\frac{1}{10}$ as non-padding frame. In inference, we fixed the length of output waveform. And we use average magnitude to distinguish the padding part. In practice, we calculate average magnitude per 1024 samples and cutoff $\le 0.02$ parts.

\section{Experiment}
\label{sec:experiment}
We compare our model with other neural TTS systems. Following \cite{chen2021wavegrad}, baseline systems were trained on a proprietary dataset consisted of 385 hours of high-quality US English speech from 84 professional voice talents. A female speaker was chosen from the training dataset for evaluation.
We implemented our model with parameter size in Table \ref{tab:params}. For pretrained BERT, we take base parameter size model trained on English only data~\footnote{English only, uncased BERT-Base model provided in \url{https://github.com/google-research/bert}}. In inference, we use 1000 steps DDPM, noise scheduling is 
\begin{equation}
\alpha_n =  \exp{(\ln(1\mathrm{e}{-7})*{(1-\cos(\frac{n}{1000}*\frac{\pi}{2}))}^{\frac{3}{2}})}
\end{equation}

Performance is measured using subjective listening tests, performed by a pool of native speakers listening with headphones.
Results are reported as the mean opinion score (MOS) which measures the naturalness of generated sample on a ten-point scale from 1 to 5.
Each sample is rated at least twice by two different native speakers.
We compared our model with character based TTS models from \cite{weiss2021wavetacotron}, the result are shown in Table \ref{tab:tts_hol}.

\begin{table}[htp]
\begin{tabular}{l|llll}
\toprule
\textbf{Block index} & 0 & 1 & 2 & 3 \\
\midrule
\textbf{Base dimension}  & \multicolumn{1}{l}{128} & \multicolumn{1}{l}{256} & \multicolumn{1}{l}{512} & \multicolumn{1}{l}{1024} \\
\textbf{Kernel Size}     & {[}5,5{]}               & {[}5,5{]}               & {[}5,5{]}               & {[}3,3,3,3,3{]}           \\
\textbf{Strides}         & {[}2,2{]}               & {[}2,2{]}               & {[}4{]}                 & {[}4,2,2,2,2{]}           \\
\textbf{Adaptive Kernel} & {[}8,8{]}               & {[}4,4{]}               & {[}2{]}                 &                           \\
\textbf{Blocks}          & {[}2,2{]}               & {[}2,2{]}               & {[}2{]}                 & {[}1,1,1,1,1{]}           \\
\textbf{Self Attention}  & {[}{\small $\times$,$\times$}{]}               & {[}{\small $\times$,$\times$}{]}               & {[}{\small $\times$}{]}                 & {[}{\small \checkmark,\checkmark,\checkmark,\checkmark,\checkmark}{]}           \\
\textbf{Cross Attention} & {[}{\small $\times$,$\times$}{]}               & {[}{\small $\times$,$\times$}{]}               & {[}{\small $\times$}{]}                 & {[}{\small \checkmark,\checkmark,\checkmark,\checkmark,\checkmark}{]}           \\
\textbf{Attention Heads} &                         &                         &                         & {[}8,8,8,8,8{]}           \\ \bottomrule
\end{tabular}
\caption{
Model configuration. Empty cell indicates it is not used in this block.
}
\label{tab:params}
\end{table}

\begin{table}
\begin{tabular}{ll}
\toprule
\textbf{Mode} & \textbf{MOS} \\
\midrule
\bfseries Ground truth & \bf{$4.56\pm0.04$} \\
\midrule
\bfseries Two-Stage Models \\
\quad Tacotron-PN + Griffin-Lim \cite{1164317} (char)  & $3.68\pm0.08$ \\
\quad Tacotron + WaveRNN \cite{pmlr-v80-kalchbrenner18a} (char) & $\bf{4.36\pm0.05}$ \\
\quad Tacotron + Flowcoder \cite{pmlr-v97-kim19b} (char) & $3.34\pm0.07$ \\
\midrule
\bfseries End-to-End Models \\
\quad Wave-Tacotron \cite{weiss2021wavetacotron} (char) & $ 4.07\pm0.06$ \\
\quad Our Model & $\bf{4.24\pm0.06}$ \\
\bottomrule
\end{tabular}
\caption{TTS performance on the proprietary single speaker dataset, evaluation contains about 600 examples.}
\label{tab:tts_hol}
\end{table}

Results suggest the proposed method leads to a better fidelity than other end-to-end systems.
One minor difference here is that the proposed system is based on sub-word instead of characters but we believe it should be comparable for TTS application.

\section{Applications}
\label{sec:application}
In this Section, we demonstrate our model could be applied in different scenarios. Specifically, we use a base model trained without any speaker information provided. The speaker is dynamically determined during inference. To enlarge the speaker diversity, we train the model on all LibriTTS data, mixing clean-100, clean-360, other-500 splits.

\subsection{Zero Shot Learning}
In the proposed approach, the alignment between the audio and text features is dynamically determined during inference.
This enables zero-shot learning for a variety of applications.
We demonstrate the model's ability through the following tasks. Examples of each task and corresponding audio samples are displayed on \url{https://e3tts.github.io}.

\subsubsection{Waveform Prompt based TTS}
For this task, for each example, we select two sentences from  same speaker from test split of LibriTTS-clean.
We concatenate the text of sentences and provide the waveform of first sentence as the prompt to the model. The prompt part are guaranteed to be longer than 3 seconds, and the part asked to be generated are guaranteed to be longer than 2 seconds, the total length are guaranteed to be shorter than 9 seconds.
During inference, on the prompt part, we replace the predicted $\epsilon$ with actual $\epsilon$. And on the rest, we keep the predicted $\epsilon$. Quantitative results are shown in the top part of Table~\ref{tab:apte}.
We report SQuId score~\cite{sellam2023squid} which is an approximation of the mean opinion score, and speaker similarity (Speaker Sim).
Speaker similarity is estimated based on the speaker embedding given by a LSTM-based speaker model~\cite{rikhye2021personalized}.
Results demonstrate that our model could generate high-quality audio given prompt with similar speaker characteristics.

\subsubsection{Text-based Speech Editing}
To evaluate the model's ability to edit speech, we evaluate the performance of text-based speech inpainting, which
is a special case of replacement. We select sentences from test split of  LibriTTS-clean and masked a small fragment (0.5 secs $\sim$ 2.5 secs) in waveform. We then provide the sentences, the masked waveform to the model, and ask it to get the whole waveform. Similar to the audio prompt task, we replace the predicted $\epsilon$ with true $\epsilon$ on the unchanged audio and keep the rest. In practice, the length of the masked part is unknown, and is usually provided by the user or predicted by some other model in a statistical way. To show the ability of diffusion model, we feed 3 example of same sentence to the model, with different masked part length($0.8\times,1.0\times,1.2\times$ the ground truth length), and reported their result in Table~\ref{tab:apte}.
From the experiment results, we can conclude that the proposed model E3 is robust against different lengths of editing span.

\begin{table}[htp]
\centering
\begin{tabular}{ll|cc}
\toprule
\textbf{Task}      & \textbf{Split} & \textbf{SQuId} & \textbf{Speaker Sim}  \\ \midrule
Prompt TTS         & Ground Truth   &       3.81         & \\
                   & Our Model      &         \bf{3.75}       & \bf{0.95} \\ \midrule
Text Editing & Ground Truth   &              3.91  & \\
                   & 0.8$\times$      &        \bf{3.84}        & 0.98\\
                   & 1.0$\times$      &        3.83        & \bf{0.98}\\
                   & 1.2$\times$      &        3.81        & 0.97\\
                   & Best of 3     &       3.85        & 0.98\\
\bottomrule
\end{tabular}
\caption{Audio prompt and text-based editing results. Metrics include SQuId which approximates mean opinion score and speaker similarity. Prompt TTS task contains about 200 examples. Text Editing task contains about 80 examples. Evaluation data are generated from test split of LibriTTS}
\label{tab:apte}
\end{table}

\subsubsection{Speaker Similarity}
For this task, we select sentence from random unseen speakers from test split of LibriTTS-clean. For each example, we select waveform $w_A$ from Speaker A and 8 waveform $w_B^1 ... w_B^8$ from random selected speakers (Speaker A must be included). We ask model to predict which speaker $w_A$ belongs to.
In inference, we concatenate the $w_A$ and $w_B^i$ and get 8 waveform. We random select a timestep ($0.04\le\alpha_n\le0.96$) and feed the noised waveform to the model.
Similar to \cite{li2023diffusion}, we calculated the L2 distance on predicted $\epsilon$ and true $\epsilon$ and sum them up using a fixed calculated weight. To make the result independent to $w_B^i$'s magnitude and length, we only take the $\epsilon$ part on $w_A$ in consideration. We summarize the result from different timestep samples using Monte Carlo method. The result for different sample times are listed in \ref{tab:similarity}.
In general, with more timestep sampled, we observe better speaker accuracy.
The result itself is interesting especially since the model is trained without any speaker information.

\begin{table}[htp]
\centering
\begin{tabular}{l|c}
\toprule
\textbf{\#Timesteps}          & \multicolumn{1}{l}{\begin{tabular}{@{}c@{}}\textbf{Speaker Classification} \\ \textbf{Accuracy}\end{tabular}} \\ \midrule
1 sample   & 75.50\%                                                  \\
4 samples  & 81.00\%                                                  \\
32 samples & \bf{83.20}\%                                                  \\ \bottomrule
\end{tabular}
\caption{Speaker similarity results evaluated on about 1000 examples. With more sampled steps, we observe better classification accuracy.}
\label{tab:similarity}
\end{table}

\subsection{Sample Diversity}

\begin{table}[htp]
\centering
\begin{tabular}{l|c}
\toprule
                       & \textbf{Fréchet Speaker Distance}   \\   \midrule
Ground Truth & 8.38 \\ \midrule
Wave-Tacotron & 26.58                                    \\
Our Model     & \bf{12.30}                                     \\ \bottomrule
\end{tabular}
\caption{Fréchet Speaker Distance results on proprietary dataset. Fréchet Speaker Distance measures the audio diversity. FSD of the ground truth is measured by computing the score between non-overlap subsets of ground truth audio.}
\label{tab:fsd}
\end{table}

Diffusion models can generate samples with higher diversity and with more authentic distribution. To measure distribution, inspired by FID~\cite{heusel2018gans} and FAD~\cite{kilgour2019frechet}, we introduce new metric called Fréchet Speaker Distance (FSD). We use a LSTM-based speaker model~\cite{rikhye2021personalized} and take the normalized last embedding layer. And calculate the distance using 
\begin{equation}
    FSD_{A, B} = {\|\mu_A - \mu_B\|} ^2+ Tr(C_A+C_B-2\sqrt{C_A*C_B})
\end{equation}
where $\mu$ represent the mean of model output speaker embedding among all examples, $C$ represent the covariance. 

We evaluate our model's FSD score on proprietary dataset.
Results in Table~\ref{tab:fsd} reveals the proposed E3 TTS system greatly improves the diversity comparing to previous work.
It reaches similar score as the ground truth.

\section{Conclusion}
\label{sec:summary}
We have proposed a novel end-to-end text-to-speech (TTS) model, E3, that is capable of generating high-fidelity audio directly from BERT features.
E3 is based on the diffusion model, an iterative refinement process.
The alignment between the audio and text features is dynamically determined during generation using cross attention.
E3 greatly simplifies the design of end-to-end TTS systems and has been shown to achieve impressive performance in experiments.
We also demonstrate that this simplified architecture enables a variety of zero-shot tasks, such as speech editing and prompt-based generation.

In future work, we plan to extend E3 to support multilingual speech generation by replacing the English-only BERT model with a multilingual language model.

\end{document}